\documentclass[twocolumn,aps,pra,amsmath,floatfix]{revtex4}
\usepackage{graphicx}
\begin{document}

\title{Absence of orbital-selective Mott transition in 
         Ca$_{2-x}$Sr$_x$RuO$_4$}
\author{A. Liebsch}
\affiliation{Institut f\"ur Festk\"orperforschung, Forschungszentrum
             J\"ulich, 52425 J\"ulich, Germany}
\date{\today }

\begin{abstract}
Quasi-particle spectra of the layer perovskite Sr$_2$RuO$_4$ are 
calculated within Dynamical Mean Field Theory for increasing values of 
the on-site Coulomb energy $U$. At small $U$ the planar geometry splits 
the $t_{2g}$ bands near $E_F$ into a wide, two-dimensional $d_{xy}$ 
band and two narrow, nearly one-dimensional $d_{xz,yz}$ bands.
At larger $U$, however, the spectral distribution of these states
exhibit similar correlation features, suggesting a common 
metal-insulator transition for all $t_{2g}$ bands at the same 
critical $U$.
\end{abstract}
\maketitle

Sr$_2$RuO$_4$ has attracted considerable interest during the recent 
years because of a variety of fascinating properties. It exhibits 
unconventional p-wave superconductivity\ \cite{maeno,phystoday}
and is in fact the only known layered perovskite without copper that 
becomes superconducting in the absence of doping. The ground state is 
non-magnetic whereas the volume perovskite SrRuO$_3$ is a strong 
ferromagnet. Substitution of Sr with Ca makes Sr$_2$RuO$_4$ undergo 
a transition to an antiferromagnetic Mott insulator\ \cite{nakatsuji,cao}. 
In contrast, doping with La leads to a ferromagnetic instability 
accompanied by a breakdown of Fermi-liquid behavior\ \cite{kiku}.  
Spin fluctuations in Sr$_2$RuO$_4$ are believed to have both 
ferromagnetic and antiferromagnetic components\ \cite{mazin}. Moreover, 
inelastic neutron scattering data indicate a soft phonon mode associated 
with in-plane rotations of oxygen octahedra\ \cite{braden}. 
Clearly Sr$_2$RuO$_4$ is close to structural and magnetic instabilities. 
Substitution of Sr by the smaller Ca ions also causes a crystallographic
distortion consisting of tilting and rotation of oxygen octahedra.
These modifications reduce the effective hopping between Ru $4d$
orbitals via O $2p$ levels. The resulting narrowing of the $4d$
bands is believed to be the driving mechanism for the metal-insulator
transition observed in  Ca$_{2-x}$Sr$_x$RuO$_4$ for $x=0.2$\
\cite{nakatsuji,cao,anisimov}. 

The layer perovskite
Sr$_2$RuO$_4$ can be regarded as a prototype of a strongly correlated, 
highly anisotropic multi-orbital transition metal oxide with coexisting 
wide and narrow bands near the Fermi level. Due to the planar crystal 
structure the partially filled $t_{2g}$ bands separate into a 
two-dimensional $d_{xy}$ band with a van Hove singularity (vHs) just 
above $E_F$\ \cite{lda},
and two nearly one-dimensional $d_{xz,yz}$ bands. In the 
single-particle picture these bands do not hybridize for symmetry reasons. 
Local electron-electron interactions in the Ru $4d$ shell, however, are 
strong and according to angle-resolved photoemission spectra\ \cite{pes}
the on-site Coulomb energy lies between the single-particle band widths 
of the $d_{xz,yz}$ and $d_{xy}$ bands: $W_{xz,yz}<U<W_{xy}$\ \cite{liebsch}.
Thus, although correlations primarily affect the narrow $d_{xz,yz}$ 
bands, multi-orbital interactions also influence the wider   
$d_{xy}$ bands, in particular its singularity above $E_F$. 
Recent quasi-particle calculations\ \cite{liebsch} based on the 
Quantum Monte Carlo (QMC)
Dynamical Mean Field Theory (DMFT)\ \cite{vollhardt,georges,pruschke}
showed that on-site correlations give rise to a charge transfer from the 
narrow bands to the wider $d_{xy}$ band and to a red-shift of the van
Hove singularity to within approximately 10 meV of $E_F$. Thus, doping
with a few \% La moves the Fermi level through the $d_{xy}$ singularity, 
thereby causing deviations from Fermi liquid behavior\ \cite{kiku}.

At the surface of Sr$_2$RuO$_4$, a $\sqrt2\times\sqrt2$ lattice
reconstruction driven 
by a soft phonon mode involving rotations of O octahedra reduces the 
effective $d-d$ hopping\ \cite{matzdorf}. As a result of the slightly 
narrower width of the $d_{xy}$ band in the first layer the van Hove 
singularity is shifted below $E_F$, converting the $\gamma$ sheet 
at the Fermi surface from electron-like to hole-like. Thus, 
photoemission spectra reveal a superposition of bulk bands with 
$E_{vHs} > E_F$ and surface bands with $E_{vHs} < E_F$\ \cite{shen}. 
The bulk bands are then consistent with the Fermi surface 
derived from de Haas-van Alphen measurements\ \cite{dHvA}.

To investigate the metal-insulator transition induced by iso-electronic 
substitution of Sr with Ca, Anisimov {\it et al.}\ \cite{anisimov} recently
performed DMFT calculations for Sr$_2$RuO$_4$ for increasing values
of the local Coulomb repulsion $U$. While in reality the transition is 
driven by a distortion of O octahedra, giving rise to a narrowing 
of the Ru $4d$ band without any significant increase in $U$, these
authors argued that a qualitative understanding of the Mott transition
can be achieved by increasing $U$ and ignoring the modifications of the 
density of states due to the structural transition. This picture is very 
interesting from a conceptual point of view since in the anisotropic 
$t_{2g}$ configuration $U$ is comparable at first to 
$W_{xz,yz}\approx 1.3$~eV and subsequently to $W_{xy}\approx3.5$~eV. 
A remarkable result of these calculations
were two successive Mott transitions for the $d_{xz,yz}$
and $d_{xy}$ bands, implying an intermediate region of $U$ or, 
equivalently, of Ca concentration, where the narrow bands exhibit an 
excitation gap while the wide band is still metallic. Only at still
larger $U$ a gap appeared for the entire  $t_{2g}$ complex.

To study this unusual electronic structure in greater detail we have 
extended our previous DMFT calculations for Sr$_2$RuO$_4$\ \cite{liebsch} 
to larger values of the on-site Coulomb energy. Surprisingly, we find a 
behavior that differs fundamentally from the one reported in Ref.\ 
\cite{anisimov}. For small $U$, the $d_{xy}$ and $d_{xz,yz}$ 
quasi-particle spectra indeed have quite different shapes owing to 
their predominantly two- versus one-dimensional electronic structure,
respectively. As $U$ increases, however, these spectral distributions 
begin to resemble one another, with only minor differences in the 
weight and shape of the coherent peak near $E_F$ and of the lower
and upper Hubbard bands. Our results therefore suggest that for 
sufficiently large $U$ the correlated electronic structure within the 
$t_{2g}$ shell dominates the one-electron hopping. As a consequence,
one common metal-insulator transition exists at an intermediate critical 
value of $U$ between those of the isolated $d_{xz,yz}$ and $d_{xy}$ bands. 
The reason for the discrepancy between the two approaches is not clear 
at present, but could be related to the fact that in Ref.\ \cite{anisimov}
the quantum impurity problem was treated within the non-crossing 
approximation (NCA)\ \cite{nca,zoelfl} whereas we use the more accurate
Quantum Monte Carlo method\ \cite{georges}. 

\begin{figure}[t]%1
  \begin{center}
  \includegraphics[width=4.5cm,height=8cm,angle=-90]{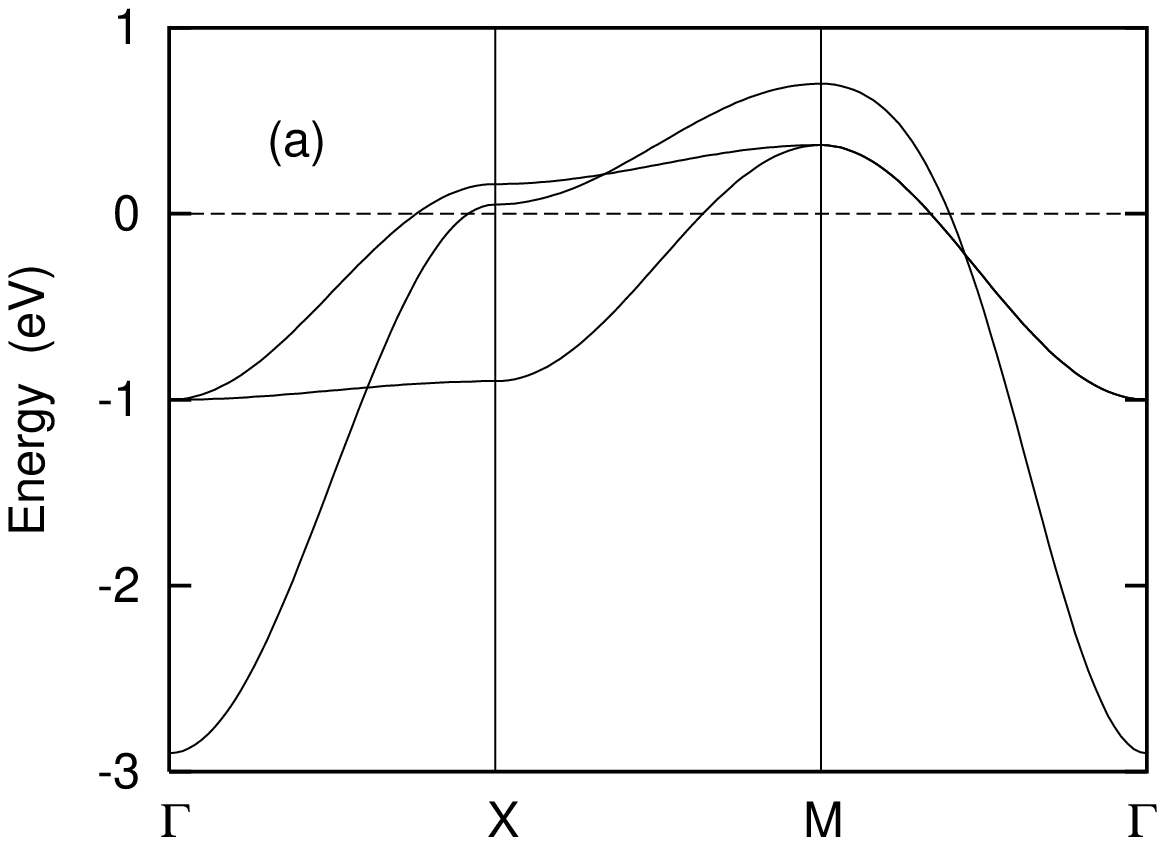}
  \includegraphics[width=4.5cm,height=8cm,angle=-90]{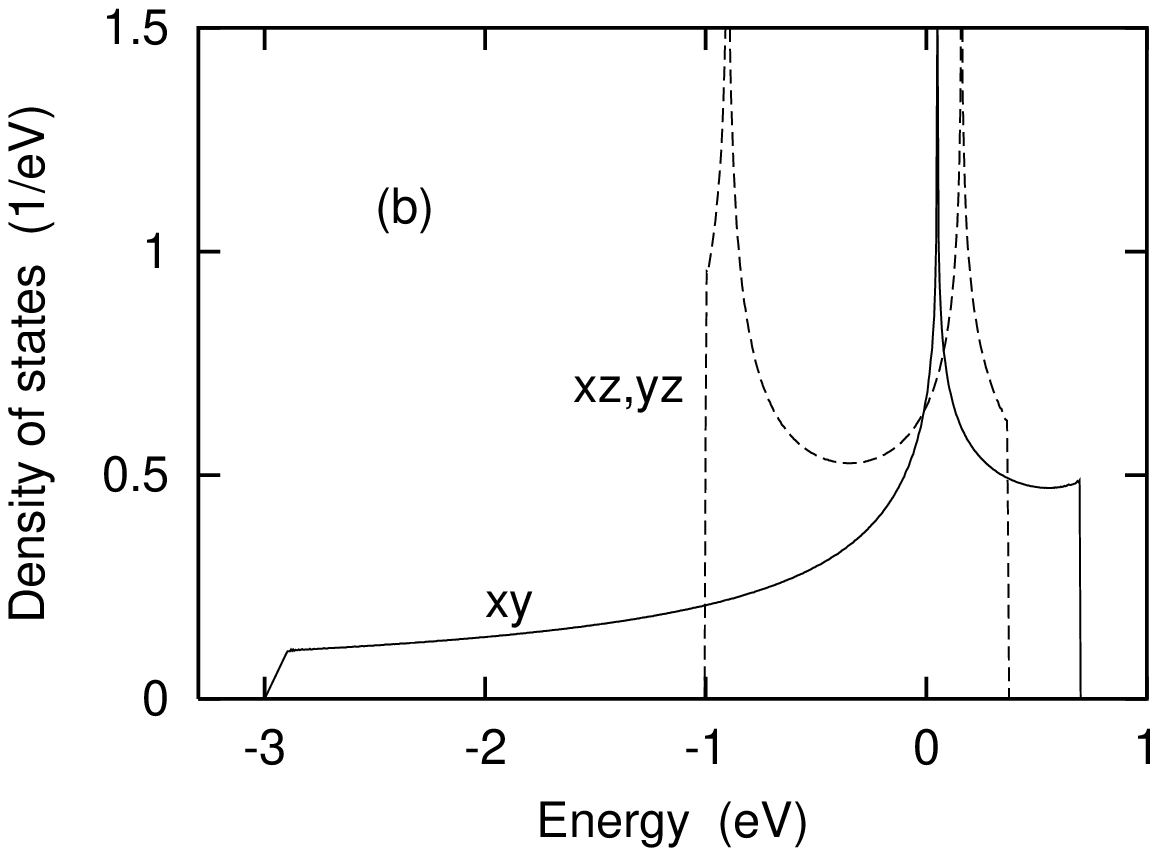}
  \end{center}
\caption{
(a) Tight-binding fit to LDA $t_{2g}$ bands of Sr$_2$RuO$_4$ ($E_F=0$); 
(b) Density of states $\rho_i(\omega)$ of $d_{xy}$ bands (solid curve) 
and of $d_{xz,yz}$ bands (dashed curve).
}\end{figure}

Fig.~1 shows the tight-binding fit to the LDA band structure of  
Sr$_2$RuO$_4$\ \cite{lda} in the vicinity of $E_F$ and the density
of states. In the bulk the $d_{xy}$ vHs is about 50~meV above $E_F$. 
All three bands are approximately 2/3 filled. 
The non-symmetric shape of the two-dimensional $\rho_{xy}(\omega)$ is 
caused by second-neighbor Ru-Ru hopping terms. The  $d_{xz,yz}$
densities are dominated by one-dimensional hopping along the Ru rows.

To take account of local Coulomb interactions we calculate the 
quasi-particle spectra within the multi-orbital QMC-DMFT approach\
\cite{liebsch,held}. Since the $d_{xy}$ and $d_{xz,yz}$ bands do not
hybridize, the self-energy is diagonal in orbital space. In the DMFT 
the elements $\Sigma_i$ ($i=xy,xz,yz$) are functionals of the bath 
Green's functions \,${\cal G}_i^{-1} = G_i^{-1} + \Sigma_i$,
where the local $G_i$ is given by   
\begin{equation}
  G_i(i\omega_n) = \int^{\infty}_{-\infty}\! d\omega\
      \frac{\rho_i(\omega)}{i\omega_n+\mu-\Sigma_i(i\omega_n)-\omega}\ 
\end{equation}
and $\mu$ is the chemical potential. Most QMC calculations were done
for $\beta=8$ with 64 time slices and several runs using $2x10^5$ 
sweeps. In the low temperature calculations for $\beta=40$ 128 time 
slices were used, with up to $5x10^4$ sweeps. The quasi-particle 
density of states \,$N_i(\omega)=-{\rm Im}\,G_i(\omega)/\pi$ is 
obtained via maximum entropy reconstruction\ \cite{jarrell}.

As shown in Ref.\ \cite{liebsch}, the angle-resolved photoemission
spectra of Sr$_2$RuO$_4$ can qualitatively be represented by quasi-%
particle distributions for local Coulomb and exchange energies 
$U=1.2$~eV, $J=0.2$~eV. The $d_{xy}$ and $d_{xz,yz}$ spectra then look 
like moderately deformed versions of their respective single-particle 
densities of states. For instance, the lower van Hove singularity of 
the $d_{xz,yz}$ bands near $-1$~eV is broadened and shifted to about 
$-0.5$~eV. Larger Coulomb energies would yield too small binding energies
for this spectral feature. This result is consistent with the fact that 
on-site Coulomb energies for $4d$ transition metals are smaller 
than for $3d$ metals. The $d_{xy}$ vHs is also shifted and lies only 
about 10~meV above $E_F$. Both bands show very low spectral weight in 
the energy region of the Hubbard peaks. The effective masses calculated 
within the QMC-DMFT for these Coulomb energies also agree with experiment.

We now increase the on-site Coulomb energy in order to explore the 
metal-insulator transition for this anisotropic multiband system.
To illustrate the effect of the different widths of the $t_{2g}$
subbands on the quasi-particle spectra we show first in Fig.~2 
the results for hypothetical three-fold degenerate bands
consisting either of $d_{xy}$ or $d_{xz,yz}$ character. The total 
filling is 4 as in actual Sr$_2$RuO$_4$. For $U=3.0$~eV the wide $d_{xy}$
band is dominated by the strong quasi-particle peak at $E_F$. Since
$U<W_{xy}$ it is only moderately affected by correlations and shows 
only weak shoulders in the range of the Hubbard bands. On the other 
hand, since $U>>W_{xz,yz}$ the narrow band derived from the $d_{xz,yz}$ 
density of states exhibits clear signs of a Mott transition. The 
upper and lower Hubbard peaks are the dominant spectral features.
Because of the finite temperature used in the QMC-DMFT calculation, 
the remaining quasi-particle peak at $E_F$ obscures the gap between the 
Hubbard bands. Nevertheless, according to the temperature dependence of 
similar features found in one-band systems\ \cite{georges,bulla} the 
value of $U$ used in the spectra shown in Fig.~2 should be very 
close to the critical value for a metal-insulator transition within 
the $d_{xz,yz}$ bands.
  
\begin{figure}[t!]%2
  \begin{center}
  \includegraphics[width=4.5cm,height=8cm,angle=-90]{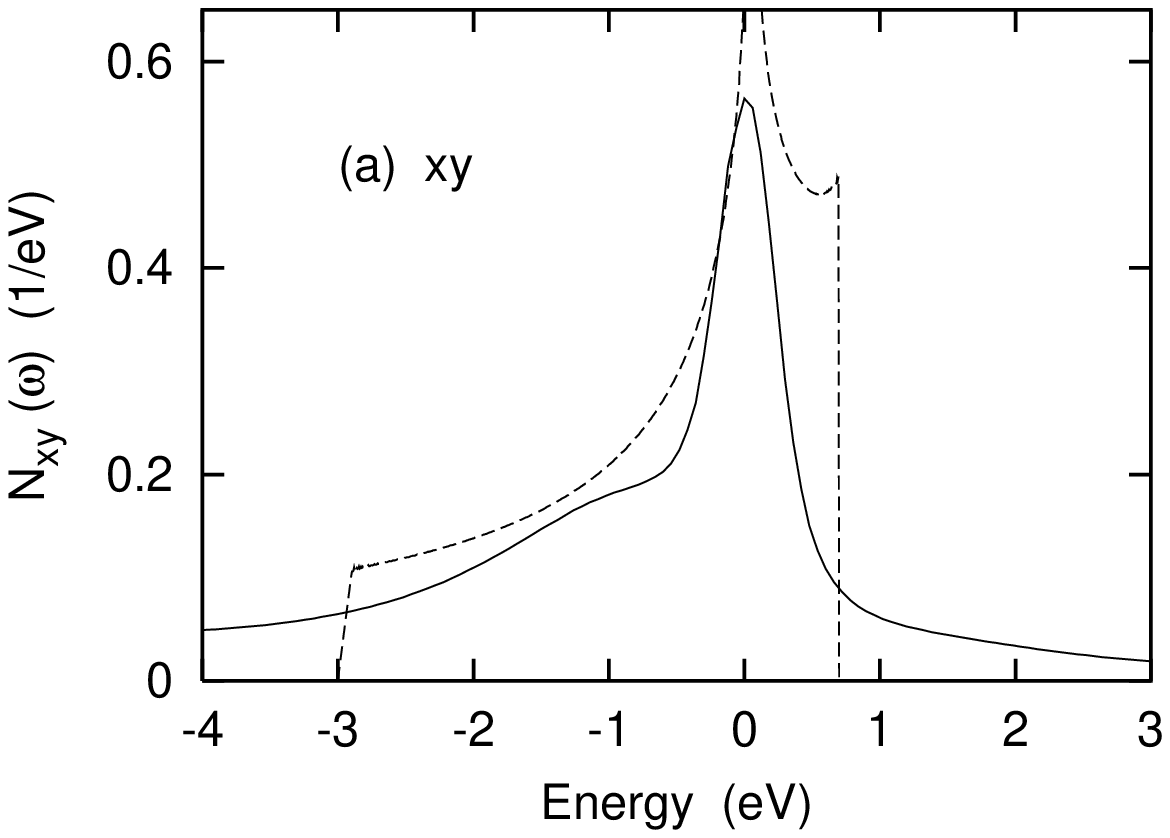}
  \includegraphics[width=4.5cm,height=8cm,angle=-90]{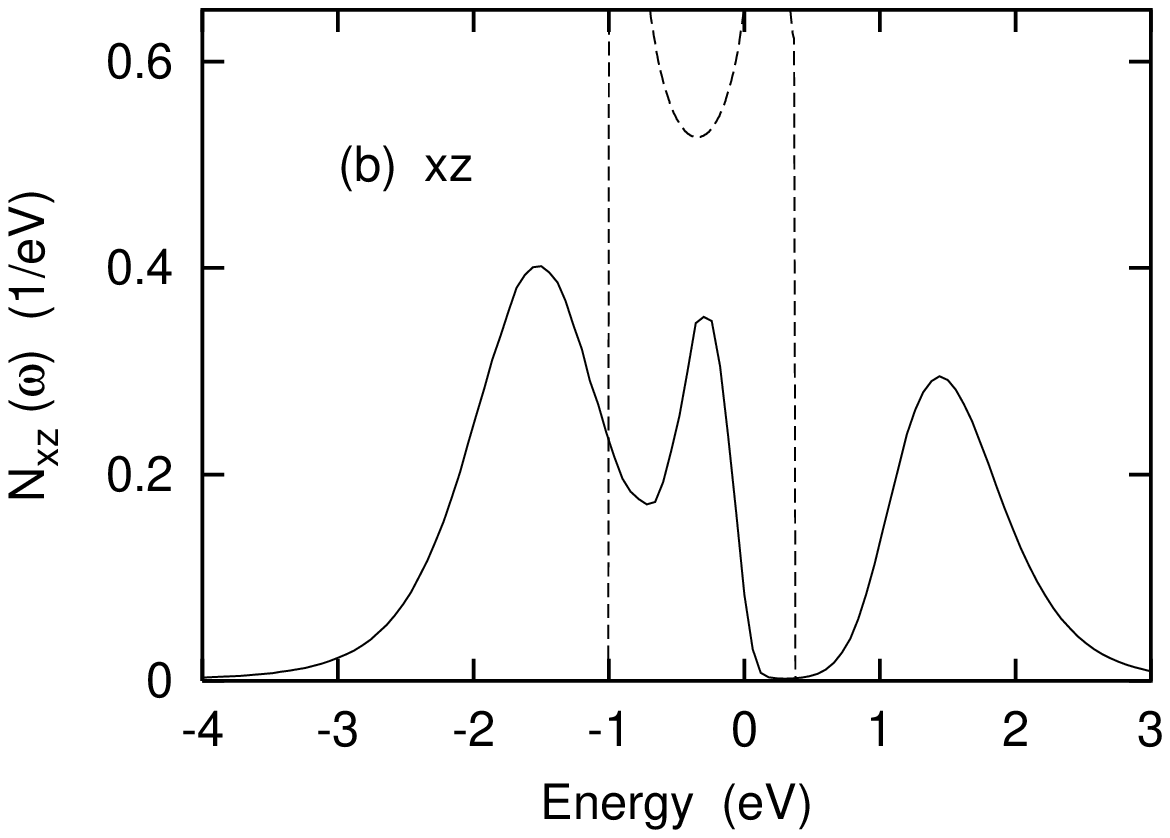}
  \end{center}
\caption{
Quasi-particle density of states $N_i(\omega)$ of Sr$_2$RuO$_4$   
derived from DMFT for $T=1450$~K, assuming independent
three-fold degenerate (a) $d_{xy}$ and (b) $d_{xz,yz}$ bands (solid 
curves; see text). $U=3.0$~eV, $J=0.2$~eV;
dashed curves: bare densities.
}\end{figure}

The quasi-particle spectra for the actual $t_{2g}$ bands of Sr$_2$RuO$_4$  
are shown in Fig.~3\,(a) for the same on-site energies. Evidently, as
as result of the strong inter-orbital Coulomb and exchange interactions
both $d_{xy}$ and $d_{xz,yz}$ spectral distributions have nearly lost 
their original density of states character and look remarkably similar:
with coherent peaks at $E_F$ and upper and lower Hubbard bands of about
the same intensity, position and shape. These spectra suggest a common
Mott transition for the $t_{2g}$ bands at a value of $U$ only slightly 
larger than 3~eV. We point out that the local correlations for $U=3$~eV
induce a charge transfer of about 0.05 electrons from the narrow 
$d_{xz,yz}$ bands to the $d_{xy}$ band. This would imply that the 
$d_{xy}$ van Hove singularity has shifted below $E_F$ and that the 
$\gamma$ sheet of the Fermi surface is hole-like rather than 
electron-like as observed in de Haas-van Alphen measurements\ 
\cite{dHvA}. 
This supports our previous choice of a smaller $U$ for the actual 
Sr$_2$RuO$_4$  $t_{2g}$ bands\ \cite{liebsch}. 

\begin{figure}[t!]%3
  \begin{center}
  \includegraphics[width=4.5cm,height=8cm,angle=-90]{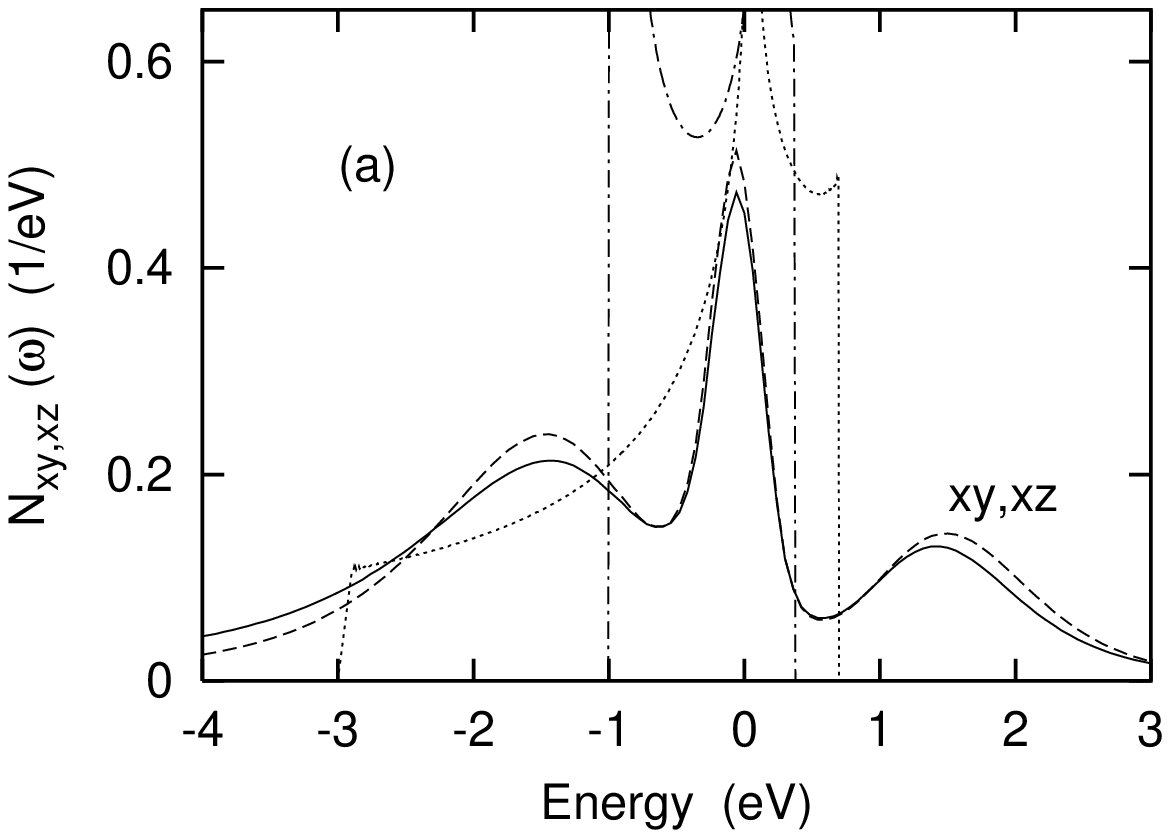}
  \includegraphics[width=4.5cm,height=8cm,angle=-90]{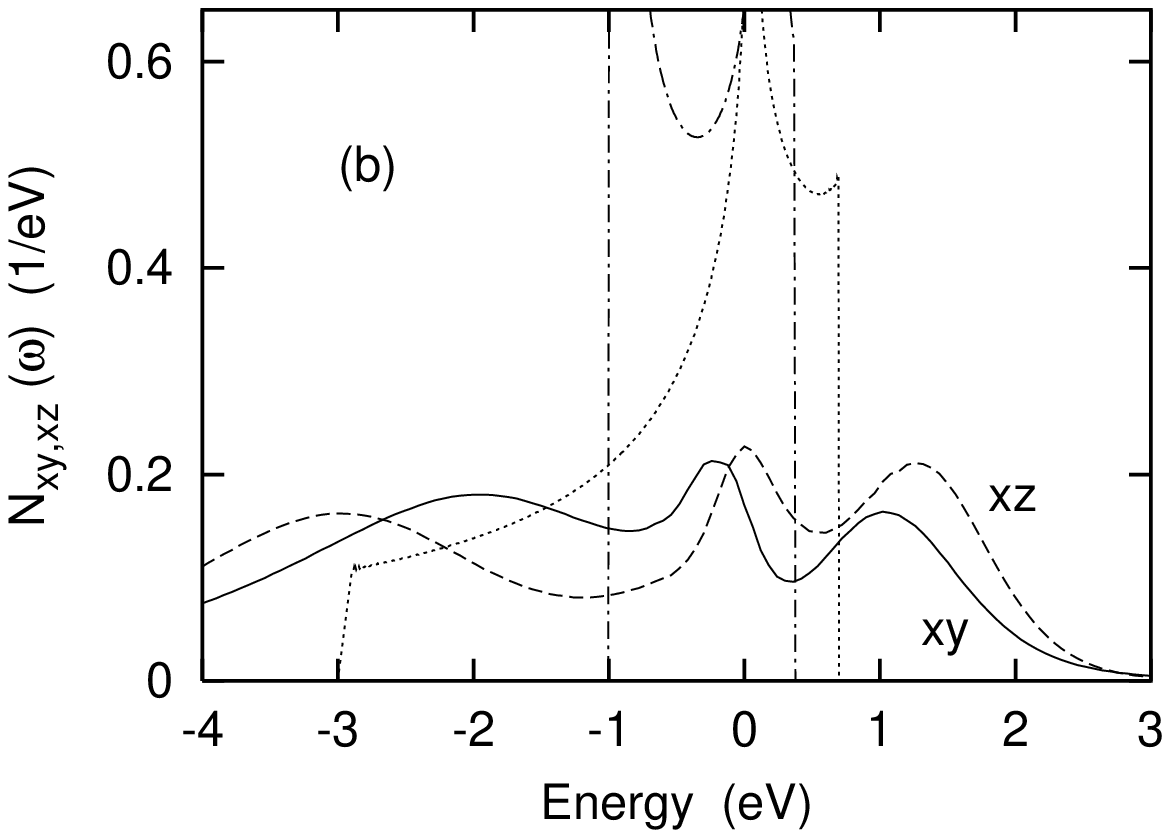}
  \end{center}
\caption{
Quasi-particle density of states $N_i(\omega)$ of Sr$_2$RuO$_4$   
derived from DMFT for $T=1450$~K. 
(a) $U=3.0$~eV, $J=0.2$~eV;
(b) $U=4.0$~eV, $J=0.7$~eV.
Solid curves: $d_{xy}$ states; dashed curves: $d_{xz,yz}$ states;
dotted and dash-dotted curves: corresponding bare densities.
}\end{figure}

Fig.~3\,(b) illustrates the effect of increasing the exchange energy
to $J=0.7$~eV. In order to keep the average Coulomb energy  $\bar U$
unchanged $U$ is increased to 4~eV. (For a $t_{2g}$ complex 
$\bar U$ coincides with the inter-orbital Coulomb energy  
$U'=U-2J$\ \cite{anisimov}. Thus, $\bar U=2.6$~eV in 
Figs.~3\,(a) and (b).) Although the  $d_{xy}$ and $d_{xz,yz}$ 
spectra are not quite as similar as in Fig.~3\,(a), they have coherent 
and incoherent peaks of about the same intensity. Only the positions 
of these features are shifted. Extrapolation to lower temperatures 
does not seem to be quite as straightforward as in the previous case. 
Nevertheless, qualitatively both  $d_{xy}$ and $d_{xz,yz}$ spectra are
equally correlated and therefore also suggest a common Mott transition.

\begin{figure}[t!]%4
  \begin{center}
  \includegraphics[width=4.5cm,height=8cm,angle=-90]{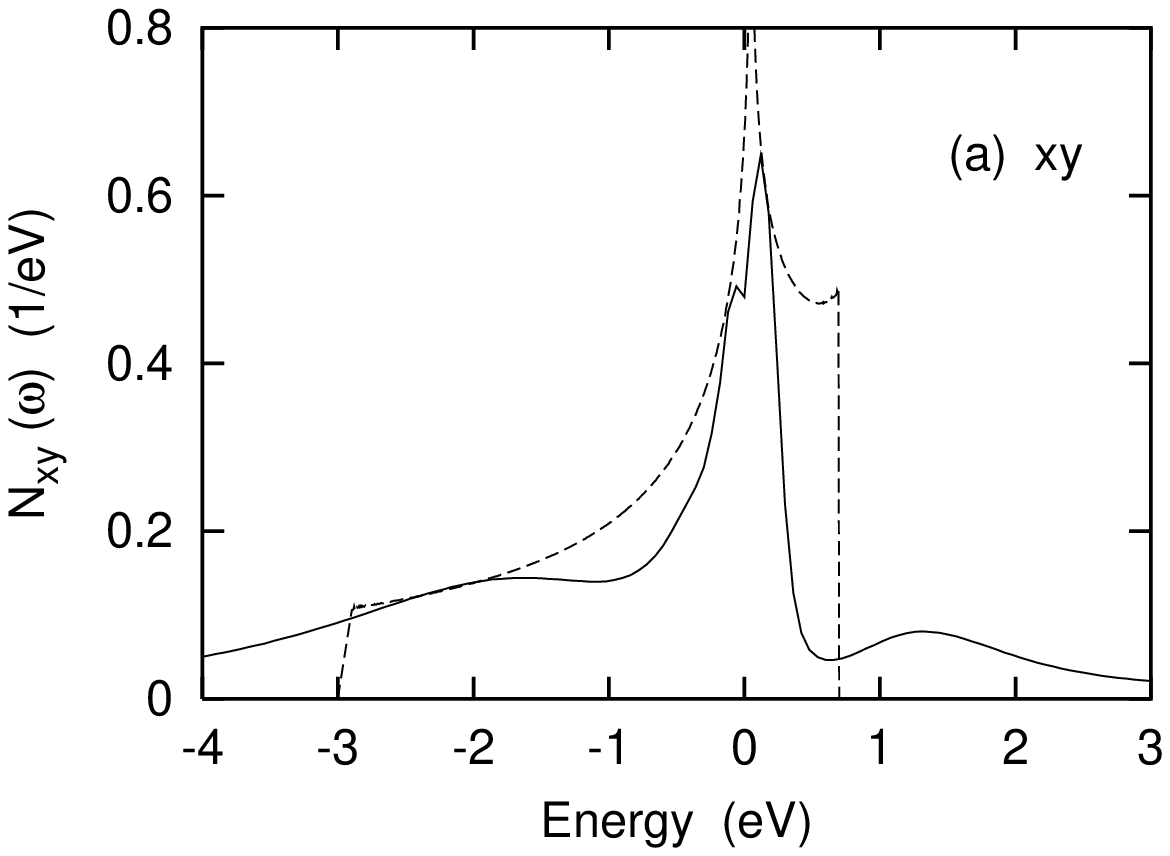}
  \includegraphics[width=4.5cm,height=8cm,angle=-90]{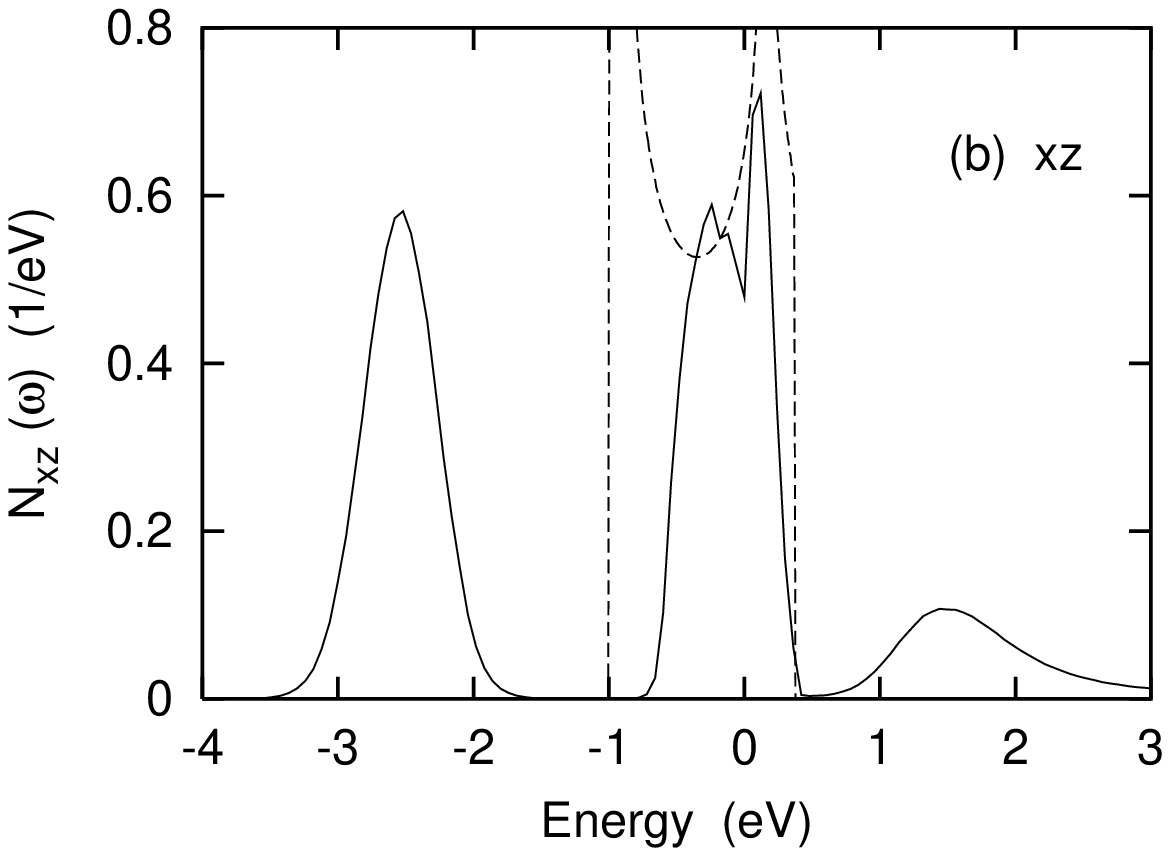}
  \end{center}
\caption{
Quasi-particle density of states $N_i(\omega)$ of Sr$_2$RuO$_4$   
derived from DMFT for $T=300$~K, $U=3.0$~eV, $J=0.2$~eV.  
Solid curves: (a) $d_{xy}$ states; (b) $d_{xz,yz}$ states;
dashed curves: bare densities.
}\end{figure}

Judging from the results discussed so far for $U$ in the critical
region it does not seem likely that upon lowering the temperature 
the $d_{xy}$ and $d_{xz,yz}$ spectra would begin to lose their 
similarity again and diverge towards two separate Mott transitions.
To test this behavior we have performed multiband QMC-DMFT calculations 
at about 300~K which are computationally much more costly.
Fig.~4 shows the $d_{xy}$ and $d_{xz,yz}$ quasi-particle distributions
for $U=3.0$~eV, $J=0.2$~eV. The peak near $E_F$ exhibits a fine structure
which is the remnant of the upper and lower edge singularities of the  
$d_{xz,yz}$ bands. The $d_{xy}$ spectrum also reveals this structure
as a result of the strong orbital interactions. The important point is
that both  $d_{xy}$ and $d_{xz,yz}$ spectra are similarly correlated:
the coherent peak has about the same weight and the upper Hubbard bands
are almost identical. The main difference is that the lower Hubbard
$d_{xz,yz}$ satellite is split off from the main peak by a  gap
while the $d_{xy}$ satellite lies in the lower tail of the $d_{xy}$
density of states and therefore forms a continuum with the main peak.
Leaving aside the uncertainties caused by the maximum entropy 
reconstruction, such a gap does not imply that the spectral weight 
at the Fermi level vanishes at a different $U_{\rm crit}$ or 
$T_{\rm crit}$ than in the case of the $d_{xy}$ states. On the contrary, 
the comparable strength of the coherent peaks indicates that the $d_{xy}$ 
and $d_{xz,yz}$ bands undergo a common Mott transition.

The above results differ qualitatively from the ones obtained by 
Anisimov {\it et al.}\ \cite{anisimov}. 
The NCA in principle should be applicable also to multiband materials.
On the other hand, even in one-band cases it is known to be unreliable
in certain situations and even violate Fermi-liquid behavior\ 
\cite{georges,held,pruschke}. It would be desirable to perform numerical 
renormalization group calculations for this system at very low 
temperatures. Unfortunately, multiband calculations within this
scheme are computationally not yet within reach.   

In summary, we have used the QMC-DMFT to calculate quasi-particle spectra 
of Sr$_2$RuO$_4$ in the range of on-site electron-electron interactions 
appropriate for a possible metal-insulator transition. Qualitatively, 
increasing $U$ might be a way to simulate the reduced $t_{2g}$ band 
width caused by distortions of O octahedra when Sr is replaced by Ca. 
At low Coulomb energies, the spectral distributions of the $d_{xy}$ and 
$d_{xz,yz}$ bands differ strongly because of the planar geometry of 
this perovskite material. Despite this anisotropy, at sufficiently 
large $U$ these spectra resemble one another in the sense that they 
are equally strongly affected by local correlations. In particular,
they exhibit similar upper and lower Hubbard bands and a coherent
peak of similar strength. These results suggest that there is a 
common Mott transition for the $t_{2g}$ complex, in contrast to 
the squential orbital-selective Mott transitions found in Ref.\ 
\cite{anisimov}. Thus, in the critical region of $U$ local Coulomb 
interactions appear to dominate the anisotropic one-electron hopping. 
       
Acknowledgement: I like to thank A. Bringer, R. Bulla, O. Gunnarsson,
A. Kampf, J. Keller, Th. Pruschke, and D. Vollhardt for useful discussions.
I also thank A.I. Lichtenstein for the QMC-DMFT code.

\end{document}